\documentclass[showpacs,preprintnumbers,amsmath,amssymb, noshowpacs]{revtex4}
\usepackage{graphicx}
\usepackage{amsmath}
\usepackage{tabularx}

\usepackage{bm}
\def\Re{\mathop{\rm Re}}
\def\Im{\mathop{\rm Im}}

\begin{document}

\title{Screening of Magnetic Moment at Co Impurity in Cu Host}

\author{A.L. Kozub$^{1,2}$, J. Koloren\u{c}$^{1}$, and A. B. Shick${}^{1}$}

\address{$^{1}$Institute of Physics ASCR, v.v.i., Na Slovance 2, Prague 8,
Czech Republic. \\
$^{2}$Faculty of Applied Physics and Mathematics, Gdansk University of Technology,  \\ Narutowicza 11/12, Gdansk, Poland.}

\begin{abstract}
Cobalt impurity located in the bulk copper is described making use of the multi-orbital Anderson impurity model that is parametrized to match the electronic structure from the local density approximation, and solved using the Lanczos method. We concentrate on the many-body description of the ground state and excitation spectra. The calculations yield a nonmagnetic ground state for the impurity atom. The computed spectral densities are in a good agreement with those obtained using the quantum Monte Carlo method.
\end{abstract}
\maketitle
\section{Introduction}

The Kondo effect, known for nearly 80 years, is one of the first discovered correlation phenomena in solid state physics~\cite{Hewson}.
It occurs when the magnetic elements with partially filled $d$-, or $f$-orbitals are incorporated into a non-magnetic host metal. 
The most prominent macroscopic hallmark of the Kondo effect  is the resistance minimum at low temperature found for metals
with magnetic impurities~\cite{Hewson}. At low temperature, the impurity spin is effectively screened by the conduction electrons. 
Although the phenomenon is well known for a long time, the theoretical description as well as the experimental investigations are 
challenging subjects of modern solid state research.

In this work we revisit the electronic and magnetic structure for the Co impurity in Cu. 
When the conventional local-spin-density approximation (LSDA), the generalized-gradient approximations (GGA)  or their Hartree-Fock-type extension (LDA+U) are used, they produce broken symmetry solutions
with  non-zero spin $M_S$ and orbital $M_L$ moments on the impurity site even without
an external magnetic field.  The true dynamical solution of an
impurity in a non-magnetic host  yields $M_S=0$ and $M_L=0$ when no external magnetic 
field is applied. In order to go beyond the static mean-field  and to incorporate the  dynamical electron
correlations, we solve a single impurity Anderson model (SIAM) whose parameters are 
extracted from LDA calculations using the finite-temperature
exact diagonalization (ED) method. 

We evaluate the spectral density at the Co impurity in Cu and 
make a comparison with published quantum Monte Carlo (QMC) results~\cite{surer2012}.
Finally, we analyse the possibility of forming a singlet ground state
which means the dynamical screening of the Co-impurity local moment by the
conduction electron bath.   

\section{LDA}
\begin{figure}[htbp]
\centerline{\includegraphics[angle=0,width=0.4\columnwidth]{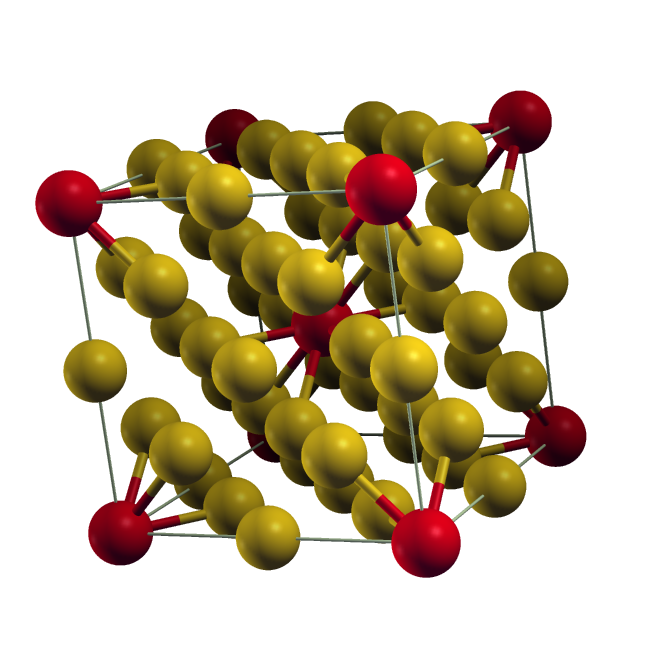}}
\caption{(Color online)The crystal structure of CoCu$_{15}$. The red spheres represent the Co atoms whereas the gold represent the Cu atoms.}
\label{fig:struct}
\end{figure}

As a computational model we use a CoCu$_{15}$ supercell 
shown in Fig.~\ref{fig:struct}. 
This supercell is chosen to keep Co and its 12 nearest Cu neighbors separated 
from other impurity atoms.  No relaxation is performed as  it is not essential for the close
packed {\em fcc} structure. We use the lattice constant of elemental Cu,
$a=6.82$~a.u.

All calculations are performed making use of a relativistic version (with spin-orbit coupling (SOC)) 
of LDA implemented in the full-potential linearized augmented plane wave  (FP-LAPW) basis~\cite{shick1997}. 
The radii of the atomic muffin-tin
(MT) spheres are set to 2.2~a.u. for both Co and Cu. The parameter $R
\times K_{\rm max}=7.7$ determined the basis set size and the
Brillouin zone was sampled with 343 $k$~points. We
checked that a finer sampling with 729 $k$~points
does not modify the results. 

\section{Single Impurity Anderson Model}
First we consider the effective SIAM
\begin{eqnarray}
\label{eq:1}
H = \sum_{k \sigma} \epsilon_{k} b_{k \sigma}^{\dagger}b_{k \sigma}  + \sum_{m \sigma}   \epsilon_{d} d_{m \sigma}^{\dagger} d_{m \sigma}
+ \sum_{m k \sigma} ( V_{mk} d^{\dagger}_{m \sigma} b_{k \sigma} + h.c.)  \nonumber \\
+ \frac{1}{2} \sum_{\substack {m m' m''\\  m''' \sigma \sigma'}}
U_{m m' m'' m'''} d^{\dagger}_{m \sigma} d^{\dagger}_{m' \sigma'} d_{m'''
\sigma'} d_{m'' \sigma}.
\end{eqnarray}
The operator $d^{\dagger}_{m \sigma}$ creates an electron in the impurity $d$ shell with energy $\epsilon_{d}$ and
$b^{\dagger}_{k\sigma}$ creates an electron in the conduction bands of the host with corresponding energy $\epsilon_{k}$. The impurity site is labeled by the magnetic quantum number $m$ and spin $\sigma=\{\uparrow, \downarrow\}$, and the conduction band of the host by a quantum number $k$ and spin $\sigma$. The conduction states are dominantly composed of $s$ and $p$ bands
of Cu. Their coupling to the impurity orbitals is described by the hybridization parameters $V_{mk}$. 
The Coulomb interaction parameters which enter the last term in Eq.(\ref{eq:1}) are taken as external parameters
of the model and their particular choice will be discussed below.  
\begin{figure}[htbp]
\centerline{\includegraphics[angle=0,width=7.0cm]{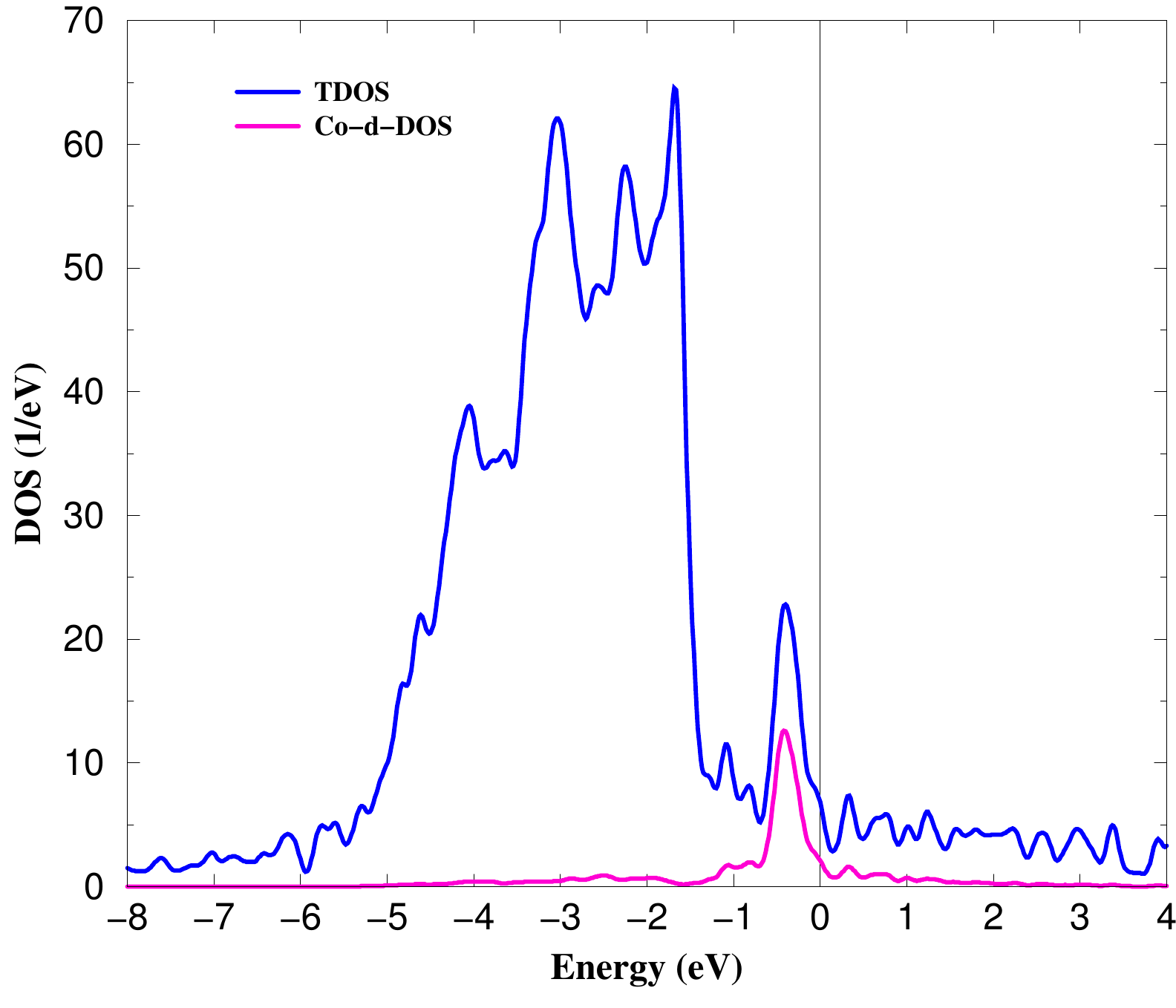}
\includegraphics[angle=0,width=7.0cm]{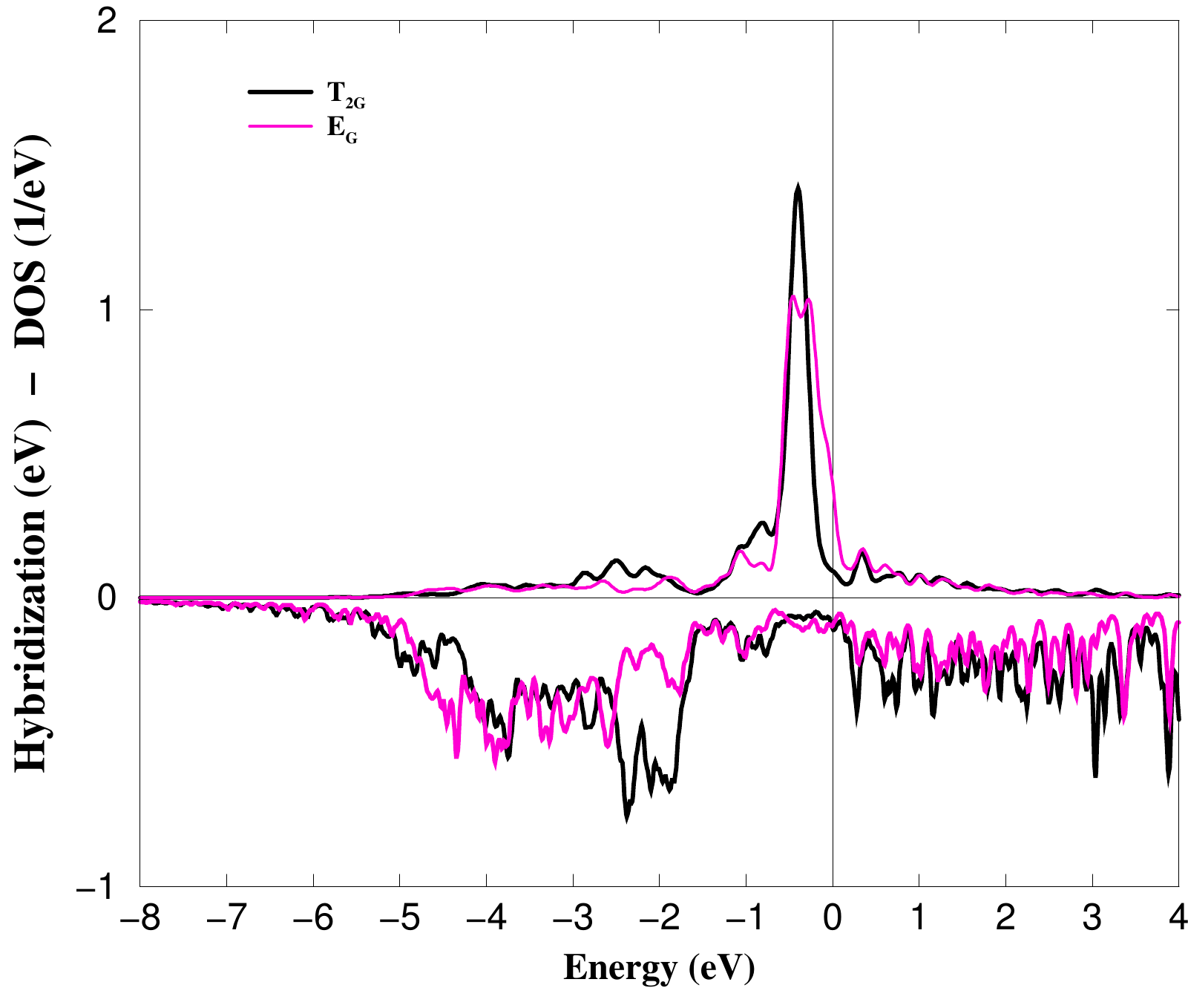}}
\caption{(Color online) (Left) The total LDA DOS (TDOS) and Co $d$-orbital LDA DOS (Co-d-DOS) of the Co impurity in Cu bulk, (right) LDA $t_{2g}$ and $e_g$ projected DOS and LDA hybridization function $\frac{1}{\pi} \Delta(\epsilon) = -\frac{1}{\pi} \Im Tr \left[G^{-1}\left(\epsilon + i \delta \right) \right]$ for  $t_{2g}$ and $e_g$ orbitals.}
 \label{fig:2}
\end{figure}

Next, following \cite{Gunnarsson89}, we  re-write
this model in terms of the energy-dependent $d$-only "bath" states
$| b (\epsilon,m,\sigma) \rangle = \frac{1}{V_{m}}  \sum_{k}  V_{mk} \delta({\epsilon - \epsilon_k})  |k \sigma \rangle$ 
where, $[V_m(\epsilon)]^2 = \sum_{k} V^2_{mk} \delta{(\epsilon -
\epsilon_k)} = -{\frac{1}{\pi}} \Delta_m(\epsilon)$.
The Hamiltonian transforms into:
\begin{eqnarray}
\label{eq:2}
H = \sum_{m \sigma} \int \epsilon  n^{b}_{m \sigma}(\epsilon) d \epsilon  + \sum_{m \sigma} \epsilon_{d} n^d_{m \sigma} 
+ \sum_{m \sigma} \int \left[ V_{m}(\epsilon) d^{\dagger}_{m \sigma} b_{m \sigma}(\epsilon) + h.c.\right] d \epsilon \nonumber \\
+ \frac{1}{2} \sum_{\substack {m m' m''\\  m''' \sigma \sigma'}}
U_{m m' m'' m'''} d^{\dagger}_{m \sigma} d^{\dagger}_{m' \sigma'} d_{m'''
\sigma'} d_{m'' \sigma},
\end{eqnarray}
where $n^{b}_{m \sigma}(\epsilon)=b^{\dagger}_{m \sigma}(\epsilon)b_{m \sigma}(\epsilon)$ and $n^d_{m \sigma} = d_{m \sigma}^{\dagger} d_{m \sigma}$ are the number operators for the bath and for the $d$ electrons of the impurity, respectively.
When the last term in Eq.(\ref{eq:2}) is omitted, the hybridisation function \mbox{$\Delta_m(\epsilon) = -\pi [V_m(\epsilon)]^2 = -\Im
[G(\epsilon)^{-1}]_m $} can be evaluated from the Green function of the  Eq.(\ref{eq:2}), and we assume that it can be approximated by
the LDA result for the Co atom in the Cu host.  For the ED method to be applicable, the continuum of the bath states
is discretized. Finally,  the complete SIAM for the five-orbital $d$ shell subject 
to the full spherically symmetric Coulomb interaction, spin-orbit coupling
and a cubic crystal field can be written as \cite{Hewson}
\begin{align}
\label{eq:hamilt}
H = & \sum_{km\sigma}
   \epsilon_{km} b^{\dagger}_{km\sigma} b_{km\sigma}
 + \sum_{m\sigma}\epsilon_d d^{\dagger}_{m \sigma}d_{m \sigma}
+ \sum_{mm'\sigma\sigma'} \bigl(\xi {\bf l}\cdot{\bf s} 
  +\mbox{\boldmath $\Delta_{\rm CF}$}
 \bigr)_{m m'}^{\sigma \; \; \sigma'}
  d_{m \sigma}^{\dagger}d_{m' \sigma'}
\\
& +  \sum_{km\sigma} \Bigl( V_{km}
  d^{\dagger}_{m\sigma} b_{km\sigma} + \text{h.c.}
  \Bigr) +
\frac{1}{2} \sum_{\substack {m m' m''\\  m''' \sigma \sigma'}} 
  U_{m m' m'' m'''} d^{\dagger}_{m\sigma} d^{\dagger}_{m' \sigma'}
  d_{m'''\sigma'} d_{m'' \sigma}.
\nonumber
\end{align}

The impurity-level position $\epsilon_d$ and the bath
energies $\epsilon_{km}$ are measured from the chemical potential $\mu$, that was set at values which yield the desired $\langle n_d \rangle$.
Parameter $\xi$ specifies the strength of the spin-orbit coupling, whereas {\boldmath $\Delta_{\rm CF}$}  corresponds to the strength of the cubic crystal field acting on the impurity. The cubic crystal field splits the $d$ orbitals of the impurity into triply-degenerate $t_{2g}=\{xy, xz, yz\}$ and doubly-degenerate $e_{g}=\{x^{2}-y^{2}, z^{2}\}$ blocks, therefore it is given by the energy differences corresponding to the orbitals $\Delta_{\rm CF}(t_{2g}) - \Delta_{\rm CF}(e_g)$. 
The parameters are determined from LDA
calculations as $\xi=0.076$~eV and $\Delta_{\rm CF}(t_{2g}) - \Delta_{\rm CF}(e_g)=0.10$~eV.

\section{Parametrization of the Impurity Model}
The Lorenzian-like shape of the LDA density of states (DOS) for Co $d$-orbitals (see Fig.~\ref{fig:2}) suggests that
it can be associated with the well-known solution of the SIAM Eq.(\ref{eq:1}) with
the constant (energy independent) hybridization function $\Delta$. Since we are interested mainly in 
the ground state and the low-energy
excitations, it seem reasonable  to take into account only bath states
near the Fermi level $E_{\rm F}$, where the main part of the LDA Co $d$-orbital DOS is located.
At first, the diagonal matrix elements of  $\Delta$ matrix  in  $\gamma=(xz,yz,xy,x^2-y^2,3z^2-r^2)$  basis
were averaged over a region $[-0.5,0.5]$ eV around $E_F$, and the $V^{k=1}_{\gamma}$ were determined as  $\sqrt{|\Delta_{\gamma}|}$.
The bath parameters $\epsilon^{k}_{\gamma}$  were chosen to reproduce the LDA $\gamma$-partial occupations
 $n^{\gamma}_d$. 
 The values of 
$V_{\gamma}^{k}$ and $\epsilon_{\gamma}^{k}$ parameters are given in Tab.~\ref{tab:1} (bath-A).
 Note that with this choice of the bath, the effective SIAM is the $d$-states charge-conserving,
 and since  the $n^{\gamma}_d$ are related to the $d$-wave phase shifts  $\delta_d^{\gamma}$ at $E_{\rm F}$
($\delta_d^{\gamma}(E_{\rm F})$=$\pi n^{\gamma}_d$),  the Friedel sum rule~\cite{Hewson} is obeyed.

The Coulomb interaction term  potential in the Eq.(\ref{eq:hamilt}) 
is given by:
\begin{eqnarray}
U_{m m' m'' m'''} = \sum_{k}a_k(m m' m'' m''')F_k, & \\ \nonumber
a_k(m m' m'' m''')    = \frac{4\pi}{2k+1} \sum_{q=-k}^{k} \langle lm|Y_{kq}|lm'' \rangle \langle lm'|Y_{kq}^{*}|lm''' \rangle,
\label{eq:4}
\end{eqnarray}
where $|lm \rangle$ is a spherical harmonic, and 
$F_k$ are the Slater integrals. The ballpark values for the
Coulomb interaction parameters $U=F_0=4$~eV and $J = 0.9$~eV ($F_2 =
7.75$~eV, $F_4= 4.85$~eV) were used in these calculations.

\begin{figure}[htbp]

\centerline{\includegraphics[angle=0,width=7.0cm]{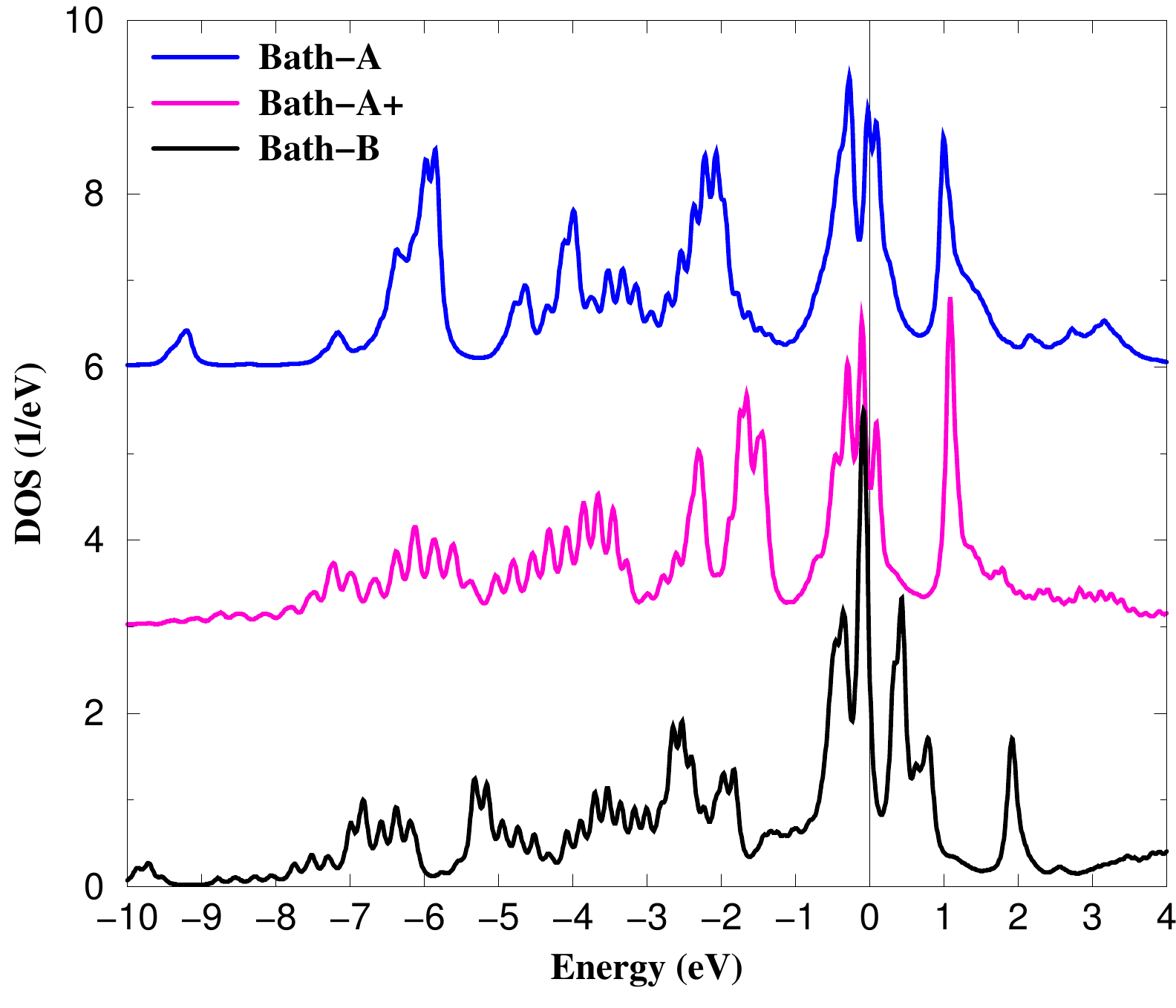}}
\caption{(Color online) The $d$-orbital spectral function of Co in bulk Cu obtained from the LDA+ED calculations for three choices of the bath. Bath-A: bath with 10 spinorbitals, Bath-A+: Bath-A extended with 10 extra spinorbital parameters $-2$~eV below the $E_F$, Bath-B: bath with 20 spinorbitals in the vicinity of $E_F$.
}
 \label{fig:3}
\end{figure}

After the parameters of the discrete impurity model are set,
the band Lanczos method~\cite{meyer89} is utilized to determine the lowest
lying eigenstates of the many-body Hamiltonian and to calculate 
one-particle Green's function $G^d_{\rm SIAM}$. The resulting
$d$-orbital spectral function $\mathop{\rm -Im}(G^d_{\rm SIAM})/\pi$
is shown in Fig.~\ref{fig:3} for the model  bath-A with $\langle n_d \rangle=7.3$,  the inverse temperature $\beta=500$
eV$^{-1}$ was used in these calculations. 

In order to examine the numerical stability of the discrete SIAM with respect to the choice
of the bath parameters, we added the extra 10 spinorbitals into the bath at the 
energy of $-2$~eV below $E_F$, as given in  Tab.~\ref{tab:1} (bath-A+) . The corresponding
$d$-orbital spectral function for the model bath-A+ is shown in Fig.~\ref{fig:3}.
Although the details of the
spectral peaks depend somewhat on the particular choice of the bath,
the overall structure of the spectrum with peak(s) in the vicinity of
the Fermi level is preserved. Also, the number of $d$-electrons $\langle n_d \rangle=7.3$ remains the same, if $\mu$ is the same.
Thus we conclude that additional orbitals away from the region
near  $E_F$ do not contribute significantly to the low energy spectrum. 

Next, we examine the effect of the extra orbitals above the Fermi level. Here, we  
take an average $\Delta_{\gamma}$  over a region $[-1,0]$ eV for the first 10 bath spinorbitals,
and over a region $[0,1]$ for another 10 bath spinorbitals. We make a symmetric choice with respect to the $E_F$ for the  $\epsilon^{k}_{\gamma}$ parameters, 
and fit the $\epsilon^{k}_{\gamma}$ parameters  to reproduce the LDA $\gamma$-partial occupations
 $n^{\gamma}_d$.
The values of $V^{k}_{\gamma}$ and $\epsilon^{k}_{\gamma}$ parameters for this "bath-B" model are given in Tab.~\ref{tab:1}.
The $d$-orbital spectral function for the model bath-B is shown in Fig.~\ref{fig:3}. It is seen that an extra (to the bath-model A) 
bath site (10 spinorbitals) modifies the spectrum in the vicinity of $E_F$ somewhat stronger than additional bath sites at 
the energies away from $E_F$. 

\begin{table*}[!ht]
\centering
\caption{Values of d-shell partial occupations $n_d$, $\Delta_{\rm CF}$ (eV)  and the bath parameters $\epsilon^{k}_{\gamma}$ (eV) and $V^{k}_{\gamma}$ (eV) 
obtained from LDA .}
\label{tab:1} 

\begin{tabular*}{\textwidth}{@{\extracolsep{\stretch{1}}} cccccc} 

\hline \hline
\multicolumn{6}{c}{Co in Cu}\\
$\gamma$& \multicolumn{1}{c}{$xz$} & \multicolumn{1}{c}{$yz$}&\multicolumn{1}{c}{$xy$}&  \multicolumn{1}{c}{$x^2-y^2$}&\multicolumn{1}{c}{$3z^2$} \\ \hline
$n^{\gamma}_d$&0.715     &0.715    &0.715 & 0.72 & 0.72 \\
$\Delta_{CF}$    &0.04   &0.04      &0.04 & -0.06 &-0.06 \\ \hline
\multicolumn{5}{c}{[5x2] bath orbitals (bath A)}\\
$\epsilon^{k=1}_{\gamma}$ & -0.025 & -0.025 & -0.025 & -0.190 & -0.190 \\ 
$V^{k=1}_{\gamma}$ & 0.385 & 0.385 & 0.385 & 0.330&0.330  \\ 
\hline
\multicolumn{5}{c}{2x[5x2] bath orbitals  (bath A+)}\\
$\epsilon^{k=1}_{\gamma}$ & -0.025 & -0.025 & -0.025 & -0.190 & -0.190 \\ 
$V^{k=1}_{\gamma}$ & 0.385 & 0.385 & 0.385 & 0.330&0.330  \\ 
\hline
$\epsilon^{k=2}_{\gamma}$ & -2.00 & -2.00 & -2.00 & -2.00 & -2.00 \\ 
$V^{k=2}_{\gamma}$ & 0.756 & 0.756 & 0.756 & 0.435&0.435  \\
 \hline
 \multicolumn{5}{c}{2x[5x2] bath orbitals  (bath B)}\\
$\epsilon^{k=1}_{\gamma}$ & -0.160 & -0.160 & -0.160 & -0.090 & -0.090 \\ 
$V^{k=1}_{\gamma}$ & 0.300 & 0.300 & 0.300 & 0.293&0.293  \\ 
\hline
$\epsilon^{k=2}_{\gamma}$ &  0.160 &  0.160 &  0.160 &  0.090 &  0.090 \\ 
$V^{k=2}_{\gamma}$ & 0.455 & 0.455 & 0.455 & 0.369&0.369  \\ 
\hline \hline
\end{tabular*}
\end{table*}
\section{Comparison with Quantum Monte Carlo}
\begin{figure}[htbp]
\centerline{
\includegraphics[width=7.0cm]{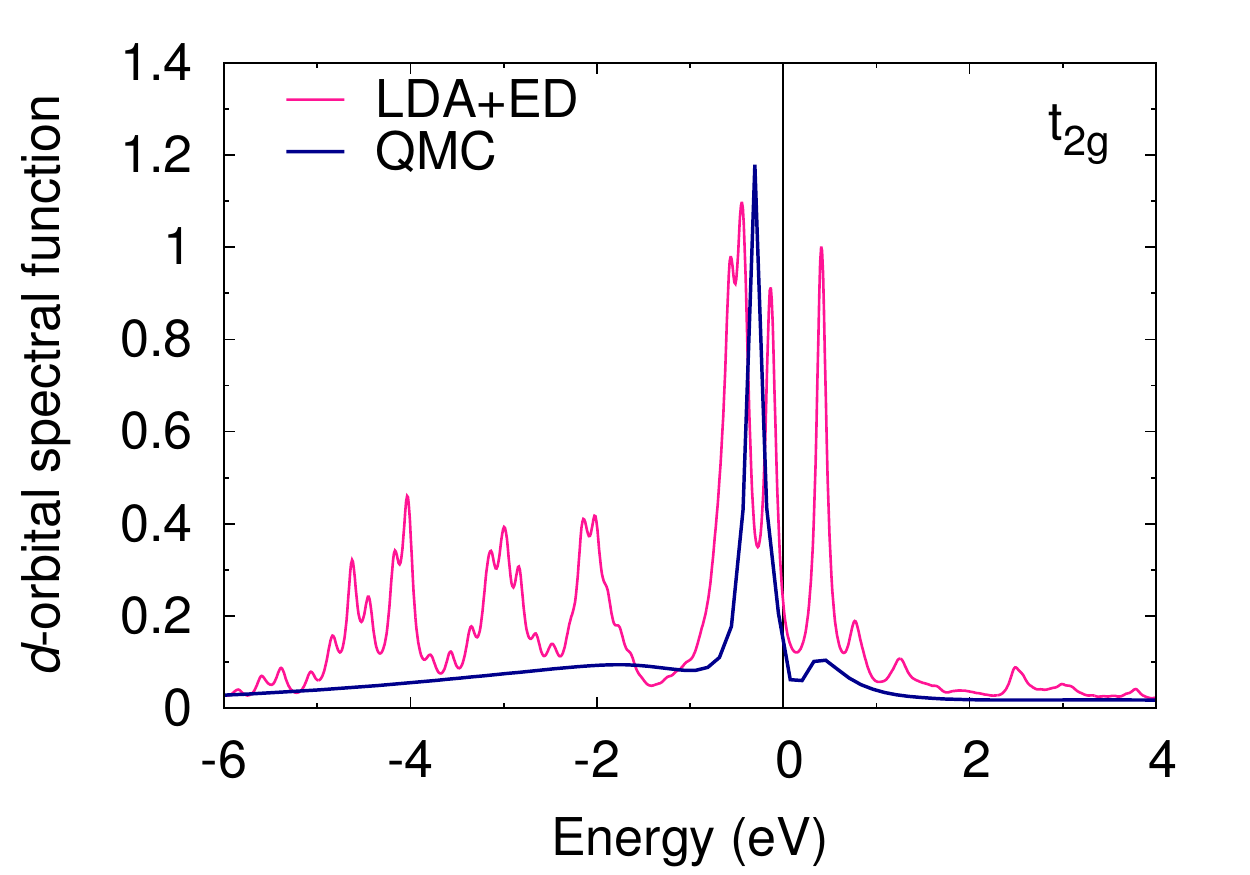}
\includegraphics[width=7.0cm]{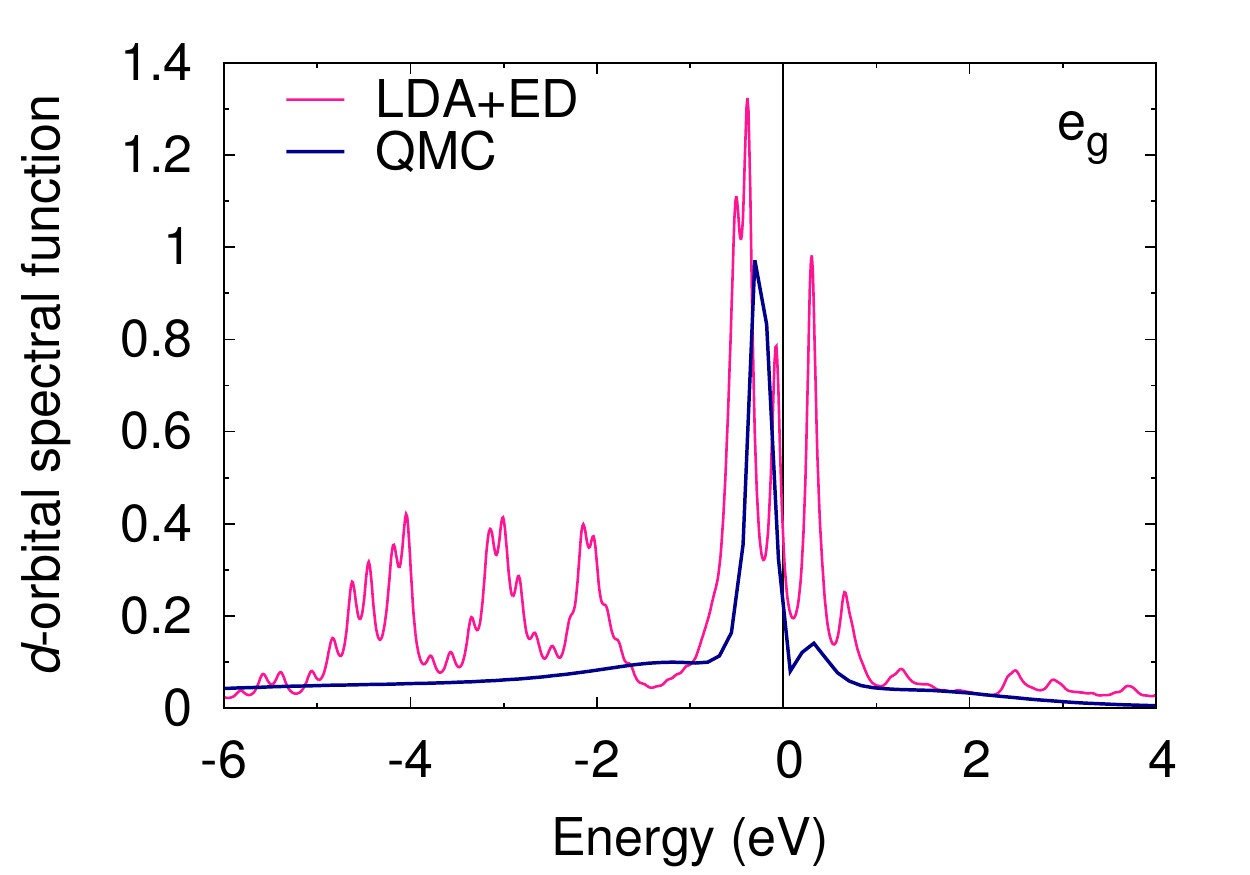}}
\caption{(Color online) The $d$-orbital spectral function for $t_{2g}$ (left) and $e_g$ (right) orbitals of Co impurity in bulk Cu for $\langle n_{d} \rangle = 7.78$. Pink 'LDA+ED' - our results, blue 'QMC' results from \cite{surer2012}.}
\label{fig:4}
\end{figure}

In order to make sure that our LDA+ED solver with discrete bath yields reasonable results, we make a comparison with continuous-time QMC calculations~\cite{surer2012} where the continuum bath is used.     
The $d$-orbital spectral function for $t_{2g}$ and $e_g$ orbitals of Co impurities in bulk Cu for $\langle n_{d} \rangle = 7.78$
are shown in Fig. \ref{fig:4} in comparison with the QMC results. In the presented LDA+ED calculations we used the bath spinorbitals from the ''bath B'' model and inverse temperature $\beta = 500$~eV$^{-1}$. 
In the vicinity of $E_F$, that is in the region where the hybridization function was fitted, the spectral density obtained from the LDA+ED calculations corresponds well to the QMC results. The single narrow peaks visible in the QMC results below $E_F$ are in the case of LDA+ED represented by three neighboring peaks in corresponding energy region. Similarly to the QMC results, the spectral function does not exhibit significant differences between $t_{2g}$ and $e_{g}$ types of orbitals.

\section{Results and discussion}
The system was analyzed for five values of the occupations of the Co $d$ shell $\langle n_{d} \rangle $ ranging between $7.2$ and $7.78$. In all considered cases the ground state of the system is a singlet, therefore the bath of conduction electrons of the host is screening the magnetic moment on the impurity site. The two highest considered occupations $\langle n_{d} \rangle = 7.51$ and $\langle n_{d} \rangle = 7.78$ correspond to the results presented also in~\cite{surer2012}.  The values of $\mu$ corresponding to those occupations are in good agreement with those from QMC calculations and are equal approximately $26\:$eV and $27\:$eV. Due to the small value of SOC, we observe only slight differences between the results
with and without it. The inclusion of the effect to the model does not change qualitatively
$d$-orbital spectral function.

Next, we calculated the spin $S$, orbital $L$ and total $J$ moments for the considered occupations from the expectation value $\langle \hat{X}^2 \rangle = X(X+1)$, where $X=S$, $L$ or $J$, respectively. The values of the moments were decreasing with increase of the occupancy $\langle n_{d} \rangle$. The spin moments lay within the range 1\,--\,1.3, the orbital moments were equal approximately 3 for all occupations and the total moments were equal 3.4\,--\,3.6. Our spin moments are slightly higher than those from QMC calculations, although still in good agreement. 
The precise values for ED SIAM, together with those for QMC are listed in Tab.~\ref{table_CoinCu}. 

Following~\cite{surer2012} we calculated orbital-resolved quasiparticle weight $Z$  for the orbitals $t_{2g}$ and $e_{g}$ according to the formula $Z_{m} = \left( 1- \frac{\Re \partial \Sigma_{m} (\omega)}{\partial \omega} \vert_{\omega \to 0}\right)^{-1}$.

The precise values of $Z(t_{2g})$ and $Z(e_{g})$ are shown in Tab.~\ref{table_CoinCu}. For both types of orbitals $Z$ increases with the increase of $\langle n_{d} \rangle$ and also exhibits good agreement with QMC results.

\begin{table*}[htbp] 
\centering
\caption{Chemical potential, occupation of the impurity $d$ shell, spin $S$, orbital $L$ and total $J$  moments, and quasiparticle weight $Z$ for both types of orbitals. Values obtained from our calculations and those presented in \cite{surer2012}. } 

\begin{tabular*}{\textwidth}{@{\extracolsep{\stretch{1}}} c c c c c c c c} \hline \hline
 & $\mu\:$(eV) &  $\langle n_{d} \rangle$ &  $ S $ & $ L $ & $ J $ & $Z(t_{2g})$ & $Z(e_{g})$     \\ \hline
LDA+ED& 24.89 &7.2 & 1.31 & 2.98 & 3.63 & 0.33 & 0.31 \\
LDA+ED& 25.26 & 7.3 & 1.26  & 2.98 & 3.61  & 0.36 & 0.34 \\
LDA+ED& 25.61  & 7.4  & 1.21 & 2.98 & 3.57  & 0.38 & 0.36 \\
LDA+ED& 26& 7.51 & 1.16 & 2.97 & 3.53  & 0.4 & 0.38 \\
LDA+ED& 26.97& 7.78 & 1.03 & 2.91 & 3.4 & 0.44 & 0.43  \\ \hline 
QMC \cite{surer2012} & 26 & 7.51 & 1.02 & -- & -- & 0.38 & 0.39  \\
QMC  \cite{surer2012} & 27 & 7.78 & 0.92 & -- & --  & 0.42 & 0.47\\ \hline \hline
\end{tabular*}

\label{table_CoinCu}
\end{table*}
\section{Conclusions}
In this work we presented the electronic and magnetic structure for Co impurity in bulk Cu calculated with the use of LDA+ED method. In our method, we solved the multi-orbital SIAM with included SOC and effect of the cubic crystal field on the Co impurity. The model was parametrized matching to the LDA electronic structure. Among the tested ways of parametrization of the bath, the most accurate choice is the bath with 20 spinorbitals whose parameters were extracted from the hybridization function in the vicinity of the Fermi level. The impurity model was then solved using the Lanczos method. 

The calculations show that the magnetic moment of the Co impurity is screened by the conduction bands of the host and as a result the ground state of the system is a nonmagnetic singlet for all analyzed average occupations of the impurity $d$ shell.
Calculated Co $d$-orbital spectral functions, spin moments and orbitally resolved quasiparticle weights are consistent with the QMC results~\cite{surer2012}.

\acknowledgments
{The support from the GACR grant No. 15-071725 is acknowledged.}


\begin{thebibliography}{99}

\bibitem{Hewson}
A.C. Hewson.
\newblock {\em The Kondo Problem to Heavy Fermions}.
\newblock Cambridge University Press, 1993.

\bibitem{surer2012}
B. Surer, M. Troyer, P. Werner, T.~O. Wehling, A.~M.
  L\"auchli, A. Wilhelm, and A.~I. Lichtenstein.
\newblock Multiorbital {K}ondo physics of {C}o in {C}u hosts.
\newblock {\em Phys. Rev. B}, 85:085114, 2012.

\bibitem{shick1997}
A.~B. Shick, D.~L. Novikov, and A.~J. Freeman.
\newblock Relativistic spin-polarized theory of magnetoelastic coupling and
  magnetic anisotropy strain dependence: Application to {C}o/{C}u(001).
\newblock {\em Phys. Rev. B}, 56:R14259--R14262, 1997.

\bibitem{Gunnarsson89}
O.~Gunnarsson, O.~K. Andersen, O.~Jepsen, and J.~Zaanen.
\newblock Density-functional calculation of the parameters in the {A}nderson
  model: Application to {M}n in {C}d{T}e.
\newblock {\em Phys. Rev. B}, 39:1708, 1989.

\bibitem{meyer89}
H.D. Meyer and S.~Pal.
\newblock A band-{L}anczos method for computing matrix elements of a resolvent.
\newblock {\em The Journal of Chemical Physics}, 91(10):6195--6204, 1989.





\end{thebibliography}
\end{document}